\documentclass[10pt,conference]{IEEEtran}
\usepackage{latexsym}
\usepackage{amsfonts}
\usepackage{amssymb}
\usepackage{amsbsy}
\usepackage{amsmath}
\usepackage[dvips]{epsfig}

\title{Spreading Information in Mobile Wireless Networks}

\author{\IEEEauthorblockN{Jinho Choi$^\dag$, Seung Min Yu$^\ddag$, and Seong-Lyun Kim$^\dag$}
\IEEEauthorblockA{$^\dag$School of Electrical and Electronic Engineering, Yonsei University\\
50 Yonsei-Ro, Seodaemun-Gu, Seoul 120-749, Korea\\
$^\ddag$Samsung Electronics, Samsung-Ro, Yeongtong-Gu, Gyeonggi-do, 443-742, Korea\\
Email: \{jhchoi, smyu, slkim\}@ramo.yonsei.ac.kr}}

\begin{document}
\maketitle

\begin{abstract}
Device-to-device (D2D) communication enables us to spread
information in the local area without infrastructure support. In
this paper, we focus on information spreading in mobile wireless
networks where all nodes move around. The source nodes deliver a
given information packet to mobile users using D2D communication as
an underlay to the cellular uplink. By stochastic geometry, we
derive the average number of nodes that have successfully received a
given information packet as a function of the transmission power and
the number of transmissions. Based on these results, we formulate a
redundancy minimization problem under the maximum transmission power
and delay constraints. By solving the problem, we provide an optimal
rule for the transmission power of the source node.
\end{abstract}


\section{Introduction}
\parskip 4pt
Near field communication (NFC) technologies enable devices in close
proximity to exchange mutual information without any infrastructure
support. Device-to-device (D2D) communication in 3GPP LTE (Long Term
Evolution) also facilitates information exchange between adjacent
devices. We call this {\it information spreading} throughout this
paper. Such information spreading via wireless networks boosts
various services, for example, mobile marketing and advertisement in
local areas \cite{Holleis10}, \cite{Ghezzi09}.

For efficient information spreading, an accurate prediction on the
number of nodes that have successfully received a given information
packet as time goes is necessary. A classical research issue in
computer science is to calculate the {\it cover time} that defines
the expected number of transmissions (or hops) until all nodes in a
given network receive a specific packet \cite {Avin05}. Applications
of the cover time analysis include searching/querying, routing,
membership services and group based communications. The cover time
analysis has been limited to the wired or the static network, though
it is extended to quantify the end-to-end delay in mobile ad hoc
networks \cite{Yu10}.

The aggregated interference analysis is necessary to calculate the
probability that a node receives a specific packet successfully. In
\cite{Haenggi09}, \cite{Andrews11}, the authors modeled wireless
networks using a stochastic point process and analyzed SIR
(signal-to-interference-ratio) distribution and outage probability.
The mutual interference between cellular users and D2D should be
considered in D2D underlaying cellular network scenario.

Some previous works dealt with the information spreading in ad hoc
networks when all nodes participate as relay nodes. The authors in
\cite{Khabbazian12} proposed a selective forwarding method based on
the minimum connected dominating set (CDS). A reliable localized
broadcast protocol using location information and acknowledgements
was proposed in \cite{Pricillar13}. In many cases, however, mobile
nodes (users) have no incentive to relay the received packet.

In this paper, we focus on the information spreading in mobile
wireless networks where all nodes move around and there is no relay.
Node mobility improves the capacity of wireless networks
\cite{Tse02}. It also brings positive effects on the information
spreading. Moving nodes can deliver information anywhere by direct
transmission. On the other hand, this may cause packet {\it delay},
which is an important parameter in the information spreading.

Another parameter is the number of {\it redundant
receptions}\footnote{The term ``redundant reception" means that a
node receives the same packet multiple times.} (i.e., waste of
resources). If the maximum transmission power is not limited, we
increase the transmission power as large as the target number of
nodes in the network can receive a given information packet at once.
In practice however, the power constraint requires multiple
transmissions when delivering the information packet to the target
nodes.

Some information spreading scenarios allow large delay. Thereby,
reducing the redundant receptions is more important than delivering
the information packet quickly. From this, we have the following
questions regarding optimal information spreading in mobile wireless
networks:
\begin{itemize}
\item How many transmissions are required for delivering a given information
packet to a certain percentage of nodes in the network?

\item What is the optimal transmission power for minimizing the total number of redundant
receptions, while keeping the delay within a reasonable level?
\end{itemize}

This paper is organized as follows. In Section \ref{sec:def}, we
describe the system model and introduce the redundancy minimization
problem. Then, we describe the mobility model and derive the average
number of MUs that have successfully received a given information
packet as a function of the transmission power and the number of
transmissions in Sections \ref{sec:homo} and \ref{sec:number}. We
solve the redundancy minimization problem and provide the optimal
transmission power and the optimal number of transmissions in
Section \ref{sec:red_min} (Proposition 4 and 5).

\section{Redundancy Minimization Problem in Information Spreading} \label{sec:def}

\vskip 5pt \normalsize Consider a cellular network composed of $N_b$
base stations (BSs), $N_u$ mobile users (MUs) and $N_s$ mobile
source nodes. The source nodes deliver a given information packet to
MUs in every $T$ second using D2D communication as an underlay to
the cellular uplink. In general, cooperative relaying like flooding
is effective for spreading information. However, overwhelming
transmission due to relaying may cause serious interference to the
cellular network. Hence, source nodes get around to impart the
information, where MUs are also moving around the entire network.
The transmission power of source nodes, ${\mu}$, is limited by
$\overline {\mu}$, and the transmission power of MU is normalized by
1.

\begin{figure}[t]
\centerline{\epsfig{figure=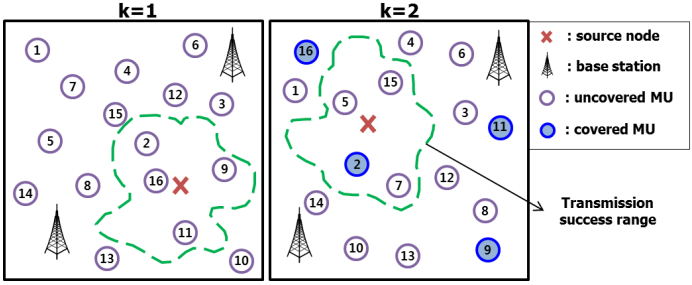,height=1.5in,clip=;}}
\caption{System model. (Numbers are inserted to discriminate the
MUs.)} \label{fig_system_model}
\end{figure}
\setlength{\textfloatsep}{10pt} \normalsize The fading between a
source node located at point $x$ and a typical MU (typical receiver)
located at the origin is ${h_x}$, and the fading between an MU who
transmits for the cellular uplink at point $y$ and the typical MU is
${g_y}$. These are assumed to be i.i.d. exponential random variables
with the unit mean (Rayleigh fading). Also, the path loss function
is given by $l(x) = {\left\| x \right\|^{ - \alpha }}$, where
$\alpha
> 2$ is the path loss exponent. For simplicity, we assume that
$\alpha = 4$. Then, for a typical mobile user, a received power of
the signal from the source node is expressed as ${\mu {h_x}{{\left\|
x \right\|}^{ - \alpha }}}$. Assuming the network is
interference-limited, the SIR (signal-to-interference-ratio) at the
typical MU is given by:
\begin{eqnarray*} SIR = \frac{{\mu
{h_x}{{\left\| x \right\|}^{ - \alpha }}}}{{\sum\limits_{y \in {\bf
\it C}} {{g_y}{{\left\| y \right\|}^{ - \alpha }}}  + \sum\limits_{z
\in {\bf \it S}\backslash \left\{ x \right\}} {\mu {h_z}{{\left\| z
\right\|}^{ - \alpha }}} }},
\end{eqnarray*}
\noindent where {\bf \it C} and {\bf \it S} denote the set of
cellular uplink MUs and source nodes, respectively. For a given
target SIR ${\beta}$, a typical MU successfully receives packets
from a corresponding source node if SIR is greater than or equal to
${\beta}$. We denote by ${p_{suc}}$ the probability that the typical
MU successfully receives packets.

\normalsize Let us define that an MU is {\it covered} if the MU
receives an information packet from a source node at least once. The
number of covered MUs by the end of the $k$-th time slot is a random
variable, denoted as $N_k$. The random variable $M_k$ denotes the
number of MUs whose SIR is not less than $\beta$ at the $k$-th time
slot, out of which $\hat M_k$ is the number of MUs that have been
already covered. For example, in the left figure of Fig.
\ref{fig_system_model}, ${M_1} = 4,\,\,\hat {{M_1}} = 0,\,{N_1} =
4$. In the right figure, ${M_2} = 5,\,\,\hat {{M_2}} = 1,{N_2} = 8$.

During the spreading process, redundant receptions may occur, which
we need to minimize as formulated below:
\setlength{\abovedisplayskip}{1pt}
\begin{eqnarray} \label{eq:formulation}
{\bf (P)} \,\,\,\,\,\, \mathop {\min }\limits_{{\mu},k} &&
f\left(\nonumber {\mu},k \right), \\ {\rm{s}}{\rm{.t}}{\rm{.}} &&
\frac{ E[ N_k ] }{N_u} \ge \gamma, \nonumber \\ && \, 0 \le {\mu}
\le \bar \mu,  \, 1 \le k \le \bar k. \nonumber
\end{eqnarray}
\noindent The objective function $f({\mu},k)$ denotes the number of
redundant receptions. Note that the control parameters are ${\mu}$
and $k$, which means that we jointly determine how large the
transmission power is set and how many times the information packet
is repeatedly transmitted. The first constraint requires that the
ratio of the covered MUs should be higher than or equal to a target
value ${\gamma}$. The second constraint determines the maximum
transmission power. The last constraint says that the number of
required transmission slots (i.e., delay) should be less than ${\bar
k}$ slots.

\section{Mobility Model: Homogeneous Condition} \label{sec:homo}
To describe node mobility, we define the {\it homogeneous condition}
\cite{Yu10} as follows:

\vskip 5pt \noindent {\bf {Definition 1}}: {\it If $E [M_k] = N_u
p_{suc}$ and $E\left[\frac{\hat M_k}{M_k}\right] = E
\left[\frac{N_{k-1}}{N_u} \right]$ for all $k$, then node mobility
is said to satisfy the homogeneous condition.}

\vskip 5pt \noindent To understand the homogeneous condition, let us
regard covered MUs as molecules of a chemical {\it solute}. Then,
the homogeneous condition resembles a homogeneous solution where the
solute concentrations in any location are the same owing to the high
speed of molecular movement. Definition 1 means that all nodes
should be uniformly distributed and the ratio of covered MUs in any
segmental area of the network should be the same with that ratio of
the whole network to satisfy the homogeneous condition. The second
figure of Figure \ref{fig_system_model} is an example satisfying the
condition. In the figure, one of the four MUs is covered in the
transmission range and four of the sixteen MUs are covered in the
whole network.

\vskip 5pt \noindent {\bf {Proposition 1}}: {\it If all nodes are
randomly distributed in the whole area and move anywhere
independently of their previous positions (i.e., the i.i.d. mobility
model \cite{Lin04}), then the network satisfies the homogeneous
condition.}

\vskip 5pt \begin{proof} If all nodes have the i.i.d. mobility, they
are uniformly distributed in the network at each time slot. The SIR
of an arbitrary MU is larger than $\beta$ with the same probability
$p_{suc}$. Thus, $M_k$ follows a binomial distribution
$B(N_u,p_{suc})$, and $E[M_k]=N_u p_{suc}$.

Moreover, the distribution of $[\hat M_k | M_k ,N_{k - 1}]$ follows
a binomial distribution $B(M_k,N_{k-1}/N_u)$, because the position
of node is independent of its previous position. Using the total
probability theorem, we calculate $E[\hat M_k / M_k]$ as follows:
\small \setlength{\abovedisplayskip}{5pt}
\setlength\arraycolsep{5pt}
\begin{eqnarray} E\left[ {\frac{{{{\hat M}_k}}}{{{M_k}}}} \right] =
{E_{{{N}_{k - 1}}}}\left[ {{E_{{M_k}}}\left[ {{E_{{{\hat
M}_k}}}\left[ {\frac{{{{\hat M}_k}}}{{{M_k}}}\left| {{M_k},{{N}_{k -
1}}} \right.}
\right]\left| {{{N}_{k - 1}}} \right.} \right]} \right] \nonumber \\
= {E_{{{N}_{k - 1}}}}\left[ {{E_{{M_k}}}\left[
{\frac{1}{{{M_k}}}\frac{{{M_k}{{N}_{k - 1}}}}{N_u}\left| {{{N}_{k -
1}}} \right.} \right]} \right] = {E_{{{N}_{k - 1}}}}\left[
{\frac{{{{N}_{k - 1}}}}{N_u}} \right]. \nonumber
\end{eqnarray}
\end{proof}
\normalsize Another mobility model that satisfies the homogeneous
condition is the random direction model \cite{Bettstetter01} with
high relative speed\footnote {By the relative speed, we mean the
moving speed relative to the transmission interval $T$.}. In the
random direction model, all nodes' speeds and moving directions are
chosen randomly and independently of other nodes. If an MU has high
relative speed that is enough to reach any point in the network
during $T$, the random direction mobility model is equivalent to the
i.i.d. mobility model and satisfies the homogeneous condition, which
we will verify by means of simulations in Figure
\ref{fig_simulation}. Hereafter, we assume that our considered
network satisfies the homogeneous condition.

\section{Number of Covered Mobile Users} \label{sec:number}
\vskip 5pt In this section, we derive the average number of covered
MUs, $E[N_k]$. We consider two transmission modes; {\it broadcast}
and {\it unicast}. In the broadcast mode, all MUs whose SIR is
higher than $\beta$ receive the information packet. In the unicast
mode, the source intends to deliver the packet to the nearest MU.

To derive the average number of covered MUs, we need to know the
successful transmission probability $p_{suc}$, for which we model
the aggregate interference by stochastic geometry and shot-noise
theory \cite{Haenggi09}, \cite{Andrews11}.

\subsection{Unicast Mode}
The average number of covered MUs in the unicast mode,
$E[N^{U}_{k}]$, is expressed in the following
proposition\footnote{In this paper, the superscripts $U$ and $B$
represent the unicast mode and the broadcast mode, respectively.}:

\vskip 5pt \noindent {\bf {Proposition 2}}: {\it In the unicast
mode, the average number of covered MUs by the end of the $k$-th
time slot is}
\begin{eqnarray*}
\,E\left[ {N_k^U} \right] &=& N_u\left[ {1 - {{\left( {1 - \frac{{{N _s}}}{{{N _u}}}{p_{i}}p_{suc}^U} \right)}^k}} \right],\,\,\,\,\,\\
{\rm{where}}\,\,\,\, p_{suc}^U &=& \frac{{{N _u}}}{{{N _u} +
\frac{\pi }{2}\frac{{\sqrt {\beta} }}{{\sqrt {{{\mu}}} }}{N_b} +
\frac{\pi }{2}{N_s}\sqrt \beta}}\,\,, \\{p_{i}} &=& 1 - \frac{{{N
_b}}}{{{N _u}}}\left( {1 - {{\left( {1 + {{3.5}^{ - 1}}{{{N _u}}
\mathord{\left/ {\vphantom {{{N _u}} {{N _c}}}} \right.
\kern-\nulldelimiterspace} {{N _b}}}} \right)}^{ - 3.5}}} \right).
\end{eqnarray*}
\begin{proof}
\normalsize We approximate  $p_{suc}^{U}$ by assuming the network as
a Poisson network, which means that BSs, MUs and the sources are
located according to independent homogeneous Poisson point processes
${{\Phi _b}}$, ${{\Phi _u}}$ and ${{\Phi _s}}$, respectively. Also,
we set the intensities as $\lambda _b={{{N_b}} \mathord{\left/
{\vphantom {{{N_b}} S}} \right. \kern-\nulldelimiterspace} S}$,
$\lambda _u={{{N_u}} \mathord{\left/ {\vphantom {{{N_b}} S}} \right.
\kern-\nulldelimiterspace} S}$ and $\lambda _s={{{N_s}}
\mathord{\left/ {\vphantom {{{N_b}} S}} \right.
\kern-\nulldelimiterspace} S}$, where $S$ denotes the area of the
network. Because the source nodes deliver the information using D2D
communication as an underlay to the cellular uplink, the
interference from cellular communications should be considered. To
model the interference, we regard the intensity of cellular uplink
MUs as the intensity of the BSs (full load). Then, we obtain
$p_{suc}^{U}$ as follows:
\small
\begin{eqnarray}\label{eq:sub1}
p_{suc}^U &=& {E_X}\left[ {\Pr \left[ {\left. {SIR \ge \beta } \right|X = x} \right]} \right] \nonumber\\
&=& \int_0^\infty  {{{\cal L}_I^U}\left( {\frac{{\beta {x^\alpha
}}}{{{{\mu}}}}} \right)2\pi {\lambda _u}x{e^{ - {\lambda _u}\pi
{x^2}}}dx,\,\,\,\,\,\,\,\,}
\end{eqnarray}
\begin{eqnarray}
%\\
{{\cal L}_I^U}\left( z \right) \nonumber &=& \exp \left( { - 2\pi
{\lambda _b}\int_0^\infty  {\left( {1 - \frac{1}{{1 + z{x^{ - \alpha
}}}}} \right)x\,dx} } \right)\nonumber
\\ &\times& \exp \left( { - 2\pi {\lambda _s}\int_y^\infty {\left(
{1 - \frac{1}{{1 + z{\mu}{y^{ - \alpha }}}}} \right)y\,dy} }
\right)\!. \label{eq:sub2}
\end{eqnarray}

\normalsize \noindent ${{\cal L}_I^U}\left( z \right)$ is the
Laplace transform of random variable $I$ where the aggregated
interference $I$ is composed of two independent terms: the
interference from the cellular uplink MUs and the interference from
the sources except the desired signal. The term $s_o$ denotes the
nearest source. From (\ref{eq:sub1}) and (\ref{eq:sub2}), the
successful transmission probability for unicast mode is expressed as
follows:
\begin{eqnarray}
p_{suc}^U = \frac{{{\lambda _u}}}{{{\lambda _u} + \frac{\pi
}{2}\frac{{\sqrt {\beta} }}{{\sqrt {{{\mu}}} }}{\lambda _b} +
\frac{\pi }{2}{\lambda _s}\sqrt \beta  }}\,.
\end{eqnarray}

Also, we need to know the idle probability of an arbitrary MU,
$p_i$, because the source cannot cover the MU if the MU communicates
with a BS. By the Proposition 2 in \cite{Yu13}, the probability
($p_i$) that a randomly chosen MU is not assigned a resource block
at a given time is expressed as follows:
\setlength{\abovedisplayskip}{5pt}
\begin{equation}
{p_{i}} = 1 - \frac{{{N _b}}}{{{N _u}}}\left( {1 - {{\left( {1 +
{{3.5}^{ - 1}}{{{N _u}} \mathord{\left/ {\vphantom {{{N _u}} {{N
_c}}}} \right. \kern-\nulldelimiterspace} {{N _b}}}} \right)}^{ -
3.5}}} \right).
\end{equation}

Then, we get the recurrence relation for $E[M^{U}_{k} - \hat
M^{U}_{k}]$ and solve it as follows: \setlength\arraycolsep{0.5pt}
\small
\begin{eqnarray} \label{eq:unicast_general} &&E\left[ {M_k^U -
\hat M_k^U} \right] \mathop = \limits^{(a)} E\left[ {M_k^U}
\right]E\left[ {1 - \frac{{\hat M_k^U}}{{M_k^U}}} \right] \nonumber
\\&& \mathop = \limits^{(b)}
\frac{N_u{\lambda_s}}{\lambda_u}{p_{i}}{p_{suc}^U}E\left[
{1\!-\!\frac{{N_{k\!-\!1}^U}}{N_u}} \right] \mathop = \limits^{(c)}
\frac{N_u{\lambda_s}}{\lambda_u}{p_{i}}{p_{suc}^U}{\left( {1\!-\!
\frac{N_u{\lambda_s}}{\lambda_u}{p_{i}}{p_{suc}^U}}
\right)^{k\!-\!1}}.
\end{eqnarray}
\normalsize In the above equation, $(a)$ follow from independency
between the number of MUs and the covered ratio, and the homogeneous
condition satisfies $(b)$. By solving the recurrence relation, we
achieve (c). \setlength{\abovedisplayskip}{5pt}
\begin{eqnarray*} \label{eq:unicast_end}
E\left[ {N^{U}_{k} } \right] = {\sum\limits_{i = 1}^k E\left[
{M_i^U\!-\!\hat M_i^U}\right] } = N_u\left[ {1\!-\!{{\left(
{1\!-\!\frac{{{\lambda _s}}}{{{\lambda _u}}}{p_{i}}p_{suc}^U}
\right)}^k}} \right].
\end{eqnarray*}
\end{proof}

Proposition 2 indicates that the transmission power of the source
has small impact on the average number of the covered MUs. Rather,
the probability that an arbitrary MU is the nearest MU from the
source, $N_s/N_u$, is a dominant factor.

\subsection{Broadcast Mode}
The average number of covered MUs in the broadcast mode,
$E[N^{B}_{k}]$, is expressed as the following proposition:

\vskip 5pt \noindent {\bf {Proposition 3}}: {\it In the broadcast
mode, the average number of covered MUs by the end of the $k$-th
time slot is}
\begin{eqnarray*}
E\left[ {N_k^B} \right] = N_u\left[ {1 - {{\left( {1 - {p_{i}}p_{suc}^B} \right)}^k}} \right],\,\,\,\,\,\,\,\,\,\,\,\,\,\,\,\,\,\,\,\,\,\,\,\,\,\,\\
{\rm{where}}\,p_{suc}^B = \frac{{{N_s}}}{{{N_s} + \frac{\pi
}{2}\frac{{\sqrt {\beta} }}{{\sqrt {{{\mu}}} }}{N_b} + {N_s}\sqrt
\beta  \left( {\frac{\pi }{2} - {{\tan }^{ - 1}}\left(
{\frac{1}{{\sqrt \beta  }}} \right)} \right)}}\,\,.
\end{eqnarray*}

\begin{proof}
Similar to the proof of Proposition 2, we have
\small
\begin{eqnarray*}
p_{suc}^B &=& \int_0^\infty  {{{\cal L}_I^B}\left( {\frac{{\beta
{x^\alpha }}}{{{{\mu}}}}} \right)2\pi {\lambda _s}x{e^{ - {\lambda
_s}\pi {x^2}}}dx,\,\,\,\,\,\,\,\,}
%\end{eqnarray*}
%\begin{eqnarray*}
\\{{\cal L}_I^B}\left( z \right) &=& \exp \left( { - 2\pi {\lambda
_b}\int_0^\infty {\left( {1 - \frac{1}{{1 + z{x^{ - \alpha }}}}}
\right)x\,dx} } \right)\\&& \times \exp \left( { - 2\pi {\lambda
_s}\int_0^\infty {\left( {1 - \frac{1}{{1 + z{\mu}{y^{ - \alpha
}}}}} \right)y\,dy} } \right).
\end{eqnarray*}

\normalsize And the successful transmission probability is expressed
as follows:
\begin{eqnarray}
p_{suc}^B = \frac{{{\lambda _s}}}{{{\lambda _s} + \frac{\pi
}{2}\frac{{\sqrt {\beta} }}{{\sqrt {{{\mu}}} }}{\lambda _b} +
{\lambda _s}\sqrt \beta  \left( {\frac{\pi }{2} - {{\tan }^{ -
1}}\left( {\frac{1}{{\sqrt \beta  }}} \right)} \right).}}
\end{eqnarray}

In broadcast mode, the probability that an arbitrary MU is the
nearest is not considered, because the source delivers the
information packet to the multiple receivers. We get the following
results by solving the same recurrence relation as
(\ref{eq:unicast_general}): \setlength{\abovedisplayskip}{5pt}
\begin{eqnarray*}
&&E\left[ {M_k^B - \hat M_k^B} \right] = {N_u}{p_i}p_{suc}^B{\left(
{1 - {N_u}{p_i}p_{suc}^B} \right)^{k - 1}}, \\&&E\left[ {N^{B}_{k} }
\right] = {\sum\limits_{i = 1}^k E\left[ {M_i^B\!-\!\hat
M_i^B}\right] } = N_u\left[ {1\!-\!{{\left( {1-{p_{i}}p_{suc}^B}
\right)}^k}} \right].
\end{eqnarray*}
\end{proof}

\begin{figure}[t]
\centerline{\epsfig{figure=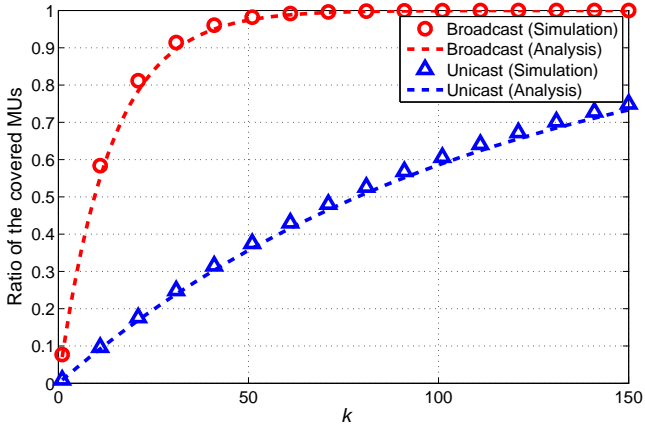,height=2.4in,clip=;}}
\caption{The average ratio of covered MUs by the end of the $k$-th
time slot in the unicast and broadcast modes. In the simulations,
all nodes move according to the random direction mobility model with
a speed of $5\,\,m/s$. $S=2000 \times 2000\,\,m^2$. ${N_b}=8$,
${N_u}= 400$, ${N_s}= 4$. ${\mu}=0.064$. $T=600s$.}
\label{fig_simulation}
\end{figure}

\vskip 5pt Different from the unicast mode, a source can deliver the
information to multiple MUs. Hence, the average number of the
covered MUs increases rapidly with transmission power. From
Propositions 2 and 3, we observe that the broadcast mode is reduced
to the unicast mode by taking $p_{suc}^B = \left( {{{{N_s}}
\mathord{\left/ {\vphantom {{{N _s}} {{N_u}}}} \right.
\kern-\nulldelimiterspace} {{N_u}}}} \right)p_{suc}^U$. Thus, we
consider only the broadcast mode in the redundancy minimization
problem (Section \ref{sec:red_min}).

To verify Propositions 2 and 3, we conducted simulations, where we
set the whole area $S=2000 \times 2000\,\,m^2$. We set the numbers
as ${N_b}=8$, ${N_u}= 400$ and ${N_s}=4$, and the transmission power
as ${\mu}=0.064$. The repeated transmission period $T=600s$. We use
the random direction mobility model \cite{Bettstetter01}, where we
set the speed of $5\,\,m/s$. MUs satisfy the homogeneous condition
because they can reach anywhere in the network in a repeated
transmission period. Figure \ref{fig_simulation} shows the results
sampled over $10^5$ instances, which exactly coincide with
Propositions 2 and 3.

\section{Optimal Rule for Redundancy Minimization} \label{sec:red_min}
\vskip 5pt In this section, we solve the redundancy minimization
problem ${\bf (P)}$. Using the fact that $\hat M^{B}_{k}$ is equal
to the number of redundant receptions caused by the transmission of
the source node at the $k$-th time slot, we derive the average
number of redundant receptions by the end of the $k$-th time slot:
\setlength{\abovedisplayskip}{5pt} \setlength\arraycolsep{1pt}
\begin{eqnarray} \label{eq:redundant_receptions}
{\it f}({\mu},k) && \mathop = \limits^{(a)} \sum\limits_{i = 1}^k
{E\left[ {M^{B}_{i} } \right]} E\left[ {\frac{{\hat M^{B}_{i}
}}{{M^{B}_{i} }}} \right] \mathop = \limits^{(b)}
N_u{p_{i}}p_{suc}^B\sum\limits_{i = 1}^k {E\left[
{\frac{{N^{B}_{i-1} }}{N_u}} \right]} \nonumber \\ &&=
N_uk{p_{i}}p_{suc}^B\!-\!N_u\left( {1\!-\!\left(
{1\!-\!{p_{i}}p_{suc}^B} \right)^k } \right)\!.
\end{eqnarray}

\normalsize \noindent where $(a)$ follows from independency between
the number of MUs and the covered ratio, and the homogeneous
condition supports $(b)$. Proposition 3 is applied to the last
equality. We can rewrite ${\bf (P)}$ as follows:
\setlength{\abovedisplayskip}{5pt}
\begin{eqnarray}
{\bf (P^\prime)} \,\,\,\,\,\, \mathop {\min }\limits_{{\mu},k} && \,\,\,\, N_uk{p_{i}}p_{suc}^B  - N_u\left( {1 - \left( {1 - {p_{i}}p_{suc}^B} \right)^k } \right), \nonumber \\
{\rm{s}}{\rm{.t}}{\rm{.}} && \,\,\,\, 1-\left( {1 -
{p_{i}}p_{suc}^B} \right)^k  \ge \gamma, \nonumber \\ && \,\,\,\, 0
\le {\mu} \le \bar \mu,\nonumber \\ && \,\,\,\, 1 \le k \le \bar k.
\nonumber
\end{eqnarray}

\normalsize Mobile devices usually have the ability of dynamically
adjusting their transmission power. Thus, we consider two cases:
$i)$ the source nodes transmit with a constant power, $ii)$ the
source nodes adjust the transmission power in every slot. For each
case, we jointly optimize ${\mu}$ and $k$ for ${\bf (P^\prime)}$.
The results are described in the following propositions.

\vskip 5pt \noindent {\bf {Proposition 4 (Optimum in constant power
case)}}:\footnote{Propositions 4 and 5 exclude the case that
$\gamma$ is equal to one because it requires the infinite number of
transmission slots.} {\it In the redundancy minimization problem
with a constant transmission power, the optimal transmission power
($\mu^*$) and the number of required transmission slots ($k^*$) are}
\begin{eqnarray*}
{\mu}^* &=& \frac{{{\pi ^2}{N_b}^2\beta}}{{4{N_s}^2{{\left( {\frac{{{p_{i}}}}{{1 - {{\left( {1 - \gamma } \right)}^{1/k*}}}} - \kappa - 1 } \right)}^2}}},\\
{k^*} &=& \left\lceil {\frac{{\log \left( {1 - \gamma }
\right)}}{{\log \left( {1 - {{{p_{i}}{N_s}} \mathord{\left/
{\vphantom {{{p_{i}}{N_s}} {\left( {{N_s}\left( {1 + \kappa }
\right) + \frac{\pi }{2}\frac{{\sqrt {\beta} }}{{\sqrt {\overline
\mu } }}{N_b}} \right)}}} \right. \kern-\nulldelimiterspace} {\left(
{{N_s}\left( {1 + \kappa } \right) + \frac{\pi }{2}\frac{{\sqrt
{\beta} }}{{\sqrt {\overline \mu } }}{N_b}} \right)}}} \right)}}}
\right\rceil,
\end{eqnarray*}

\normalsize \noindent {\it where $\lceil x \rceil$ denotes the
smallest integer that is larger than or equal to $x$ and $\kappa =
\sqrt \beta \left( {{\pi \mathord{\left/ {\vphantom {\pi  2}}
\right. \kern-\nulldelimiterspace} 2} - {{\tan }^{ - 1}}\left(
{{\beta ^{ - (1/2)}}} \right)} \right)$.}

\begin{proof}
In ${\bf (P^\prime)}$, the first constraint should be satisfied with
equality because there is no reason to cover more MUs than the
target. Therefore, we get the following equation:
\begin{eqnarray} \label{eq:relationship_k_R}
1-\left( {1 - {p_{i}}p_{suc}} \right)^k  = \gamma \,\, \to \,\,
{p_{i}}p_{suc}^{*} = \left( {1 - (1-\gamma)^{1/k} } \right).
\end{eqnarray}

\normalsize \noindent Moreover, the objective function can be
transformed into as follows: \setlength{\abovedisplayskip}{5pt}
\begin{eqnarray*}
\mathop {\min }\limits_{{\mu},k} && \,\,\,\,
N_uk{p_{i}}p_{suc}^B\!-\!N_u\left( {1\!-\!\left(
{1\!-\!{p_{i}}p_{suc}^B} \right)^k } \right)\\ \to \,\,\,\,\mathop
{\min }\limits_{{\mu},k} && \,\,\,\, N_uk{p_{i}}\left( {1 -
(1-\gamma)^{1/k} } \right)\!-\!N_u{\gamma}.
\end{eqnarray*}

\noindent The second term of the objective function in ${\bf
(P^\prime)}$ becomes $N_u\gamma$, which is independent of the
control variables ${\mu}$ and $k$. Then, we have only to minimize
the first term $N_uk{p_{i}}\left( {1 - (1-\gamma)^{1/k} } \right)$
by Equation (\ref{eq:relationship_k_R}). Note that
$N_uk{p_{i}}\left( {1 - (1-\gamma)^{1/k} } \right)$ is an increasing
function of $k$ for $0< \gamma \le 1$. Therefore, $k^*$ should be
the smallest integer that satisfies the second constraint in ${\bf
(P^\prime)}$. Using this and Equation (\ref{eq:relationship_k_R}),
we can calculate $k^*$ and ${\mu}^*$.
\end{proof}

\vskip 3pt According to Proposition 4, if the maximum transmission
power is sufficiently large or is not limited, then the optimal rule
is to increase the transmission power so large as to cover the
target number of MUs at once. On the other hand, if the maximum
transmission power is limited, ${\mu}^*$ is the largest one below
the maximum transmission power $\bar \mu$, which makes corresponding
$k^*$ be the smallest integer.

\vskip 3pt \noindent {\bf {Proposition 5 (Optimum in dynamic power
control case)}}: {\it In the redundancy minimization problem with
dynamic power control, the optimal transmission power ($\mu^*$) and
the number of required transmission slots ($k^*$) are}
\begin{eqnarray*}
{\mu}^* &=& \left\{ {\begin{array}{*{20}{c}}
{\overline \mu }&\,\,\,\,{for\,k < {k^*}}\\
{\frac{{{\pi ^2}{N_b}^2\beta{{\left[ {{{\left( {1 - {p_{i}}\overline
{{p_{suc}}} } \right)}^{{k^*} - 1}} + \gamma  - 1}
\right]}^2}}}{{4{N_s}^2{{\left[ {\left( {1 - \gamma } \right)\left(
{1 + \kappa } \right) + \left( {1 + \kappa  - {p_{i}}}
\right){{\left( {1 - \overline {{p_{suc}}} } \right)}^{{k^*} - 1}}}
\right]}^2}}}}&\,\,\,\,{for\,k = {k^*}}
\end{array}} \right.,\\
{k^*} &=& \left\lceil {\frac{{\log \left( {1 - \gamma }
\right)}}{{\log \left( {1 - {p_{i}}\overline {{p_{suc}}} }
\right)}}} \right\rceil,
\end{eqnarray*}
{\it where $\,\overline {{p_{suc}}}  = {{{N _s}} \mathord{\left/
{\vphantom {{{N_s}} {\left( {{N_s}\left( {1 + \kappa } \right) +
\frac{\pi }{2}\frac{{\sqrt {\beta } }}{{\sqrt {\overline \mu }
}}{N_b}} \right)}}} \right. \kern-\nulldelimiterspace} {\left(
{{N_s}\left( {1 + \kappa } \right) + \frac{\pi }{2}\frac{{\sqrt
{\beta} }}{{\sqrt {\overline \mu } }}{N_b}} \right)}}$}.\normalsize

\setlength{\belowdisplayskip}{4pt}
\setlength{\belowdisplayshortskip}{4pt}
\setlength{\abovedisplayskip}{4pt}
\setlength{\abovedisplayshortskip}{4pt}

\begin{proof}
Let ${\mu_{t}}$ and $R_t$ denote the transmission power of the
source node and the covered ratio of the network at $t$-th time
slot, respectively. Considering $t$-th and $(t+1)$-th time slots, it
is obvious that ${R_{t + 1}}$ is larger or equal than ${R_t}$. Then,
we can get the following equation:
\begin{eqnarray*}
{p_{suc}}\left( {{\mu _t}} \right){N_u}\left( {1 - {R_t}} \right) =
{p_{suc}}\left( {{\mu _{t + 1}}} \right){N_u}\left( {1 - {R_{t +
1}}} \right)
\end{eqnarray*}
\noindent \vskip -5pt where ${p_{suc}}\left( {{\mu _t}} \right)$
denotes the successful transmission probability which corresponds to
${\mu_{t}}$. The left-hand side of the equation means the number of
covered MUs at $t$-th time slot, and the right-hand means the number
of covered MUs at $(t+1)$-th time slot. Hence, the equation shows
the relationship among the transmission powers and the covered
ratios to cover the same number of MUs in each of two consecutive
time slots. We can rearrange the equation as
\begin{eqnarray*}
\frac{{{p_{suc}}\left( {{\mu _{t + 1}}} \right)}}{{{p_{suc}}\left(
{{\mu _t}} \right)}} = \frac{{1 - {R_t}}}{{1 - {R_{t + 1}}}}.
\end{eqnarray*}
\noindent \vskip -5pt It is obvious that right-hand side of the
equation is not less than 1. Hence, ${\mu_{t+1}}$ is larger or equal
than ${\mu_{t}}$ to satisfy the equality, because the successful
transmission probability is an increasing function of ${\mu_{t}}$.
Furthermore, we can obtain the following relationship:
\begin{eqnarray*}
{p_{suc}}\left( {{\mu _t}} \right){N_u}{R_t}  \le {p_{suc}}\left(
{{\mu _{t + 1}}} \right){N_u}{R_{t + 1}}.
\end{eqnarray*}
\noindent \vskip -5pt It means that the redundancy must be not less
than that of the previous time slot to cover the same number of MUs
at a certain time slot. With a given target number of covered MUs,
therefore, the maximum power is optimal for the redundancy
minimization except the last time slot. At the last time slot, the
transmission power that achieves the target coverage ratio is
optimal, which can be obtained from ${X_{{k^*}}} = \gamma - \left[
{1 - {{\left( {1 - {p_{i}}\overline {{p_{suc}}} } \right)}^{k^* -
1}}} \right]$.
\end{proof}

\vskip 0pt In the information spreading with dynamic power control,
the number of required transmission slots is the same as that of the
constant power case. The optimal power ${\mu}^*$ is $\bar \mu$
except the last slot in which the power that equals the average
ratio of covered MUs to ${\gamma}$.

\section{Conclusions} \label{sec:conclu}
\vskip 0pt We focused on the redundancy minimization problem for
information spreading in mobile wireless networks. By stochastic
geometry, we derived the probability that the source node
successfully delivers a given information packet to the mobile user
where the mutual interference between device-to-device communication
and the cellular communication exist. Using this, we derived the
average number of covered MUs as a function of the transmission
power and the number of transmissions for two cases; unicast and
broadcast. In unicast mode, the probability that an arbitrary
selected MU is the nearest and have not been covered is a dominant
factor for receiving the information. Hence, an algorithm for
selecting uncovered MU is important to design the unicast
information spreading system. The received signal to interference
ratio is more important in broadcast mode. Hence, interference
management schemes are more important in broadcast system.

In addition, we provided the optimal transmission power and the
optimal number of transmissions in two cases: the sources transmit
with a constant power and the sources can adjust the power in every
time slot. If the source nodes cannot adjust their transmission
power due to the simplicity of the device, then the maximum power is
the optimal to minimize redundant receptions. Maximal power
transmission is also optimal, even though the sources are able to
adjust transmission power in every time slot. In this case, however,
the sources reduce the transmission power in the last time slot, not
to exceed the target coverage ratio.

\section*{Acknowledgment}
\vskip 0pt This research was supported by the International Research
\& Development Program of the National Research Foundation of Korea
(NRF) funded by the Ministry of Science, ICT \& Future Planning of
Korea(Grant number: 2012K1A3A1A26034281)

\end{document}